\documentclass[aps,showpacs,prl,twocolumn,groupedaddress]{revtex4}
\usepackage{graphicx}

\newcommand{\be}{\begin{equation}}
\newcommand{\ee}{\end{equation}}

\newcommand{\cE}{\mathcal{E}}

\begin{document}

\title{Stable Patterns of Membrane Domains at Corrugated Substrates}

\author{Bartosz R\'{o}\.{z}ycki, Thomas R. Weikl and Reinhard Lipowsky}

\affiliation{Max Planck Institute of Colloids and Interfaces, Science Park Golm, 14424 Potsdam, Germany}

\date{\today}

\begin{abstract}
Multi-component membranes such as ternary mixtures of   lipids and cholesterol can exhibit coexistence regions between two liquid phases. When such 
membranes adhere to a corrugated substrate, the phase separation process
strongly depends on the interplay between substrate topography, bending rigidities, and  line tension  of  the membrane
domains  as we show theoretically via energy minimization and Monte Carlo simulations. 
For sufficiently large bending rigidity contrast between the two membrane phases, 
the corrugated substrate truncates the phase separation process and leads to a stable 
pattern of membrane domains. 
Our theory is consistent with recent experimental observations and provides a possible control mechanism  for domain patterns  in biological membranes.
\end{abstract}


\maketitle

{\sl Introduction} -- Biomimetic and biological membranes are 2-dimensional liquids,   
 in which the lipid molecules undergo 
fast lateral diffusion.  In general, the interactions between 
the different lipid species   may lead to the formation of intramembrane domains with  
distinct lipid composition. For giant vesicles prepared from  three-component membranes, such domains can be directly observed 
by optical microscopy  \cite{diet01,baum03+baum05,veat03+veat05,baci05,lipo278}. 
These experiments have confirmed the process of domain-induced budding 
as  predicted theoretically  \cite{lipo99+lipo104}.  The latter shape transformation provides direct
evidence for the line tension of 
the domain boundaries, which   was found to vary 
between  10$^{-12}$ and   10$^{-14}$ N for different compositions and temperatures 
\cite{baum03+baum05} and 
must vanish at the  critical demixing or consolute point of the membrane mixture \cite{lipo99+lipo104}.

The three-component membranes studied in  
\cite{diet01,baum03+baum05,veat03+veat05,baci05,lipo278}
consisted of  a saturated lipid such as sphingomyelin, 
an unsaturated phospholipid, and 
cholesterol. The same lipid mixtures were
proposed to form domains in cell membranes  \cite{simo97+schuck04,mukh04} but direct
imaging of these latter domains or `rafts' has turned out to be difficult. In contrast, it is now 
well established that the three-component model membranes exhibit
 two-phase coexistence regions in which  the membrane components separate into  liquid-ordered (${\cal L}_o$) and   liquid-disordered (${\cal L}_d$)  domains  that  are enriched in the saturated and unsaturated lipids, respectively.  

\begin{figure}[b]
\begin{center}
\resizebox{0.75\columnwidth}{!}{\includegraphics{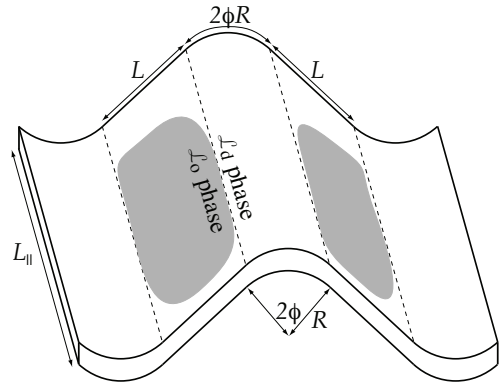}}
\caption{Multi-component membrane adhering to  a corrugated substrate surface --  
In this example, the topography is periodic in one direction and  characterized  by  the curvature radii $\pm R$ of the cylindrical ridges and valleys, the width $L$ of the flat surface segments,  and the tilt angle $\phi$. The membrane contains 
liquid ordered   (${\cal L}_o$, gray) and liquid disordered (${\cal L}_d$, white) domains. The ${\cal L}_o$ phase, which has a higher bending rigidity  than the ${\cal L}_d$ phase, tends to avoid the curved membrane regions  along the ridges and valleys.  }
\label{cartoon}
\vspace*{-0.3cm}
\end{center}
\end{figure}

The micrometer-size domains observed in three-component membranes must arise from the 
growth and coarsening of much smaller domains, a process which acts to reduce the line energy of the 
domain boundaries between the ${\cal L}_o$ and ${\cal L}_d$ domains. 
This coarsening process can be modified by  constraints on the membrane 
shape as  observed in recent experiments   \cite{Yoon06, Parthasarathy06}, in which   the
  membranes  adhered to a substrate surface.  
For a  planar substrate, the membrane domains should again undergo the same coarsening process as for giant vesicles. For
a corrugated substrate surface as  in Fig.~\ref{cartoon}, on the other hand, 
the domains tend to form certain  domain patterns as observed in  \cite{Yoon06, Parthasarathy06}. 
One important and open question  is whether these domain patterns are metastable  or represent the 
true equilibrium states of the membranes. 

 In this letter, we will address and answer this question. First, we will clarify  the basic 
 mechanism underlying the observed domain patterns  and show that these 
patterns  arise from the  
competition between the line tension, $\lambda$,  of the domain boundaries and the bending 
rigidities 
 $ \kappa_{o}$ and $ \kappa_{d}$ of the  ordered  and disordered liquid phases, where we take
 $ \kappa_{o} >  \kappa_{d}$ as    in \cite{baum03+baum05}.  
The bending 
 rigidity contrast $ \Delta \kappa \equiv \kappa_{o} - \kappa_{d}$ 
 and the line tension $\lambda$ define the elastic length scale $  \xi_{\rm el} \equiv  \Delta \kappa / \lambda$
 which must be compared with another length scale, $\xi_{\rm to}$, that is determined by the surface
 topography alone. 
 If the elastic length scale   $\xi_{\rm el}$ exceeds the topographical length scale $\xi_{\rm to}$, 
the pattern with  many striped domains is predicted to be globally stable and to  represent  the  true equilibrium 
state of the membrane. For $\xi_{\rm el} < \xi_{\rm to}$, on the other hand, the  equilibrium state 
consists of one large ${\cal L}_o$ and one large ${\cal L}_d$  domain,  both of which cover many ridges and valleys of the surface topography. 
\cite{footnote2}

Our  letter is organized as follows. We first define our theoretical model in terms of  the different energies of a multi-component membrane that forms two types of intramembrane domains and adheres to a topographically structured surface. 
Using this model, we then determine the phase diagram both for the grand-canonical 
ensemble,  in which the membrane is in contact with reservoirs for the different 
lipid molecules, and for the canonical ensemble, in which the membrane has a constant 
number of lipid molecules. Both phase diagrams are quantitatively  confirmed by extensive Monte Carlo 
simulations,  which provide additional insight into the  anisotropic coarsening process of the membrane domains.

{\sl Theoretical Model} -- The system under consideration is  a multicomponent lipid membrane,  
which strongly adheres to a topographically structured substrate as in Fig.~\ref{cartoon}. 
If the membrane is quenched into (or prepared within) the  ${\cal L}_o$--${\cal L}_d$
coexistence region, it forms two types of membrane domains corresponding to these two phases
and characterized by bending rigidities $\kappa_{o}$ and $\kappa_{d}$.  
The bending energy of the membrane is then given by 
$\cE_{\rm be}  = \int_{{\cal L}_o}  {\rm d} A \, 2 \kappa_{o}  M^2  +  \int_{{\cal L}_d}
 {\rm d} A \,  2 \kappa_{d} M^2$ where $M$ denotes the mean curvature of the  adhering  membrane and 
 the two surface integrals extend over the total area of the  ${\cal L}_o$ and ${\cal L}_d$  domains, respectively.   

The energy of the domain boundary separating the two lipid phases in the membrane is $\cE_{\rm li}  = \int_{\partial {\cal L}_o}  {\rm d} l \, \lambda  $, which depends on the line tension  $\lambda >0$   and the total length  $\int_{\partial {\cal L}_o} {\rm d} l$ of the domain boundaries. The competition of bending energy $\cE_{\rm be} $ and line tension energy $\cE_{\rm li} $ was first  studied in  \cite{lipo99+lipo104} for multi-domain vesicles. For adhering membranes as considered here, we must also include
 the 
adhesion energy of the membrane which has the form    $\cE_{\rm ad} = -  \int_{{\cal L}_o} {\rm d} A 
W_{o} -  \int_{{\cal L}_d} {\rm d} A W_{d}$, where $W_{o} > 0$ and $W_{d} > 0 $ denote the adhesion energy per unit area of the ${\cal L}_o$ and ${\cal L}_d$ membrane segments, respectively.  Combining all three energy contributions, the total  energy of the membrane
conformation can be written as 
\be 
\cE =  \int_{{\cal L}_o} {\rm d} A \, \left( 2 \, \Delta \kappa \, M^2  - \Delta W \right)  + \int_{\partial {\cal L}_o}  {\rm d} l \, \lambda  
\label{TotalEnergy}
\ee
with the bending rigidity contrast $\Delta \kappa = \kappa_{o} - \kappa_{d}$ 
and the adhesion energy contrast $  \Delta W \equiv W_{o} - W_{d}$. 
 Note that $\Delta W > 0$ corresponds to a substrate that attracts the 
ordered ${\cal L}_o$ phase more strongly than the disordered ${\cal L}_d$ phase. 

In the following, we will focus on the 
surface topography displayed in  Fig.~\ref{cartoon}. In this case,  the expression (\ref{TotalEnergy}) for the total energy simplifies since the 
mean curvature $M$ assumes only three distinct values: $M= 1/(2R)$ along the cylindrical ridges, $M= -1/(2R)$ along the cylindrical valleys, and  $M=0$ along the planar surface segments.

{\sl Phase diagram: Grand-canonical ensemble} -- From the theoretical point of view, 
it is convenient to first  consider the grand-canonical ensemble,  in which  the membrane is coupled to  reservoirs for the different lipid molecules and  the area fractions of the two types of membrane domains can adjust freely. In this case, minimization of the  
 total energy $\cE$ as given by (\ref{TotalEnergy}) leads to three different membrane phases. In phases (I) and (II), 
the whole membrane segment that adheres to the substrate is in the ${\cal L}_d$ phase and 
in the ${\cal L}_o$ phase, respectively. In phase (III), on the other hand, 
the two phases separate in such a way that the membrane stripes that adhere to the flat pieces of the substrate are in the more rigid ${\cal L}_o$ phase whereas the membrane parts that adhere to the ridges and valleys of the substrate stay in the ${\cal L}_d$ phase.
More precisely, 
the membrane attains phase (I) if
\begin{equation}
\frac{ \phi \, \Delta \kappa}{ \lambda \, R} > \left( 1+ \frac{2 \phi R}{L} \right) \frac{L \, \Delta W}{\lambda}  \quad {\rm and} \quad \frac{L \, \Delta W}{\lambda} < 2,
\label{regime_i}
\end{equation}
phase (II) if
\be
\frac{\phi \, \Delta \kappa}{\lambda \, R} < \min \left[ \left( 1+ \frac{2 \phi R} {L} \right) \frac{L \, \Delta W}{\lambda} , \; 2 + \frac{2 \phi R}{L}  \frac{L \, \Delta W}{\lambda} \right]
\label{regime_ii}
\ee
and phase (III) if
\begin{equation}
\frac{\phi \, \Delta \kappa}{\lambda \, R} > 2+ \frac{2 \phi R}{L} \, \frac{L \, \Delta W}{\lambda} \quad {\rm and} \quad \frac{L \, \Delta W}{\lambda} > 2. 
\label{regime_iii}
\end{equation}

\begin{figure}[h]
\begin{center}
\resizebox{0.9\columnwidth}{!}{\includegraphics{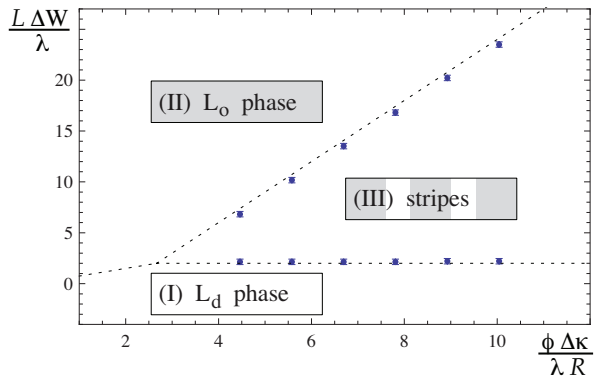}}
\caption{Phase diagram for the grand-canonical ensemble as a function of the two dimensionless parameters $\phi \Delta \kappa / \lambda R$ and $L \Delta W / \lambda$ for 
$2 \phi R /L =1/3$. 
The dashed phase boundaries have been obtained  by minimization of  the total  membrane energy $\cE$ as given by (\ref{TotalEnergy}),  the data points  by Monte Carlo simulations. Regimes (I) and (II) correspond to uniform  membranes in the ${\cal L}_d$ and ${\cal L}_o$ phase, respectively. Regime (III) is determined by the inequalities (\ref{regime_iii}) and corresponds to a striped membrane in which the membrane domains that adhere to the flat pieces of the substrate are in the more rigid ${\cal L}_o$ phase while the domains that adhere to the ridges and valleys   stay in the ${\cal L}_d$ phase. }
\label{pd3}
\vspace*{-0.3cm}
\end{center}
\end{figure}

Inspection of the  inequalities (\ref{regime_i})--(\ref{regime_iii}) shows that the overall phase 
diagram is determined by  three dimensionless parameters as given by  $\phi \Delta \kappa / \lambda R$,  $L \Delta W / \lambda$, and 
$2 \phi R /L$. The first two parameters are proportional to the bending rigidity contrast 
$\Delta \kappa$ and to 
the adhesion energy contrast $  \Delta W$,  respectively, and inversely proportional to the 
line tension $\lambda$. The third parameter $2 \phi R /L$ depends only on the topography of 
the substrate surface, compare Fig.~\ref{cartoon}. The phase diagram corresponding to  fixed 
topography with $2 \phi R /L = 1/3$ is displayed in  Fig.~\ref{pd3}. 

The three regimes in Fig.~\ref{pd3}  may be distinguished by the  area fraction $X_o$ of the ${\cal L}_o$ domains. This fraction is defined via $X_o \equiv A_o/ (A_o + A_d)$,  where $A_o$ and $A_d$ represent the total areas of the ${\cal L}_o$ and ${\cal L}_d$ domains. In regimes (I), (II), and (III), 
one has $X_o = 0$, $X_o = 1$, and $X_o = 1/(1 + 2 \phi R /L)$, respectively, which implies first
order transitions between these regimes. The three phase boundaries, 
which correspond to the dashed lines in Fig.~\ref{pd3}, 
meet in a triple point. This point is located at $2 \lambda = L \, \Delta W$ and $2 \lambda \left( 1+ 2 \phi R/ L \right) = \phi \, \Delta \kappa /R$.

\begin{figure}[h]
\begin{center}
\resizebox{0.95\columnwidth}{!}{\includegraphics{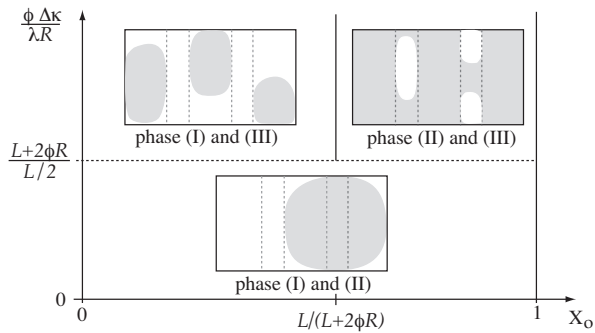}}
\caption{Phase diagram in the canonical ensemble as a function of area fraction $X_o$ and 
dimensionless rigidity contrast $\phi \Delta \kappa / \lambda R$.   The three phases (I), (II), and (III)
are now represented by three vertical lines with  $X_{o}=0$,  $X_{o}=1$ and $X_{o}=L/(2 \phi R+L)$, 
respectively.  For intermediate values of the area fraction $X_o$, one has  coexistence 
regions in which two of the three phases  coexist.  The horizontal dotted line with 
$\phi \Delta \kappa / \lambda R = 2 (2 \phi R + L)/ L$ represents transitions from a membrane with many striped domains to a membrane with two large domains. }
\label{pd_X}
\vspace*{-0.3cm}
\end{center}
\end{figure}

{\sl Phase diagram: Canonical ensemble} -- Next, let us 
 consider an adhering membrane with a certain,  fixed lipid composition. In equilibrium, the 
 area fraction $X_o$ of the ${\cal L}_o$ domains then attains a fixed value as well, and the adhesion energy contrast $\Delta W$ now plays  the role of a Lagrange multiplier in (\ref{TotalEnergy}). 
The corresponding phase diagram is shown in Fig.~\ref{pd_X} as a function of  area fraction $X_o$
 and  dimensionless rigidity contrast  $\phi \Delta \kappa / \lambda R$. The three regimes   (I) , (II), and (III) of the grand-canonical phase diagram in Fig.~\ref{pd3} are   mapped onto   three vertical lines with constant $X_o$ in the canonical phase diagram, see Fig.~\ref{pd_X}. The coexistence lines in the grand-canonical phase diagram, on the other hand, are now mapped into  three coexistence regions of  the canonical phase diagram as shown in  Fig.~\ref{pd_X}. 

\begin{figure}[h]
\begin{center}
\resizebox{0.8\columnwidth}{!}{\includegraphics{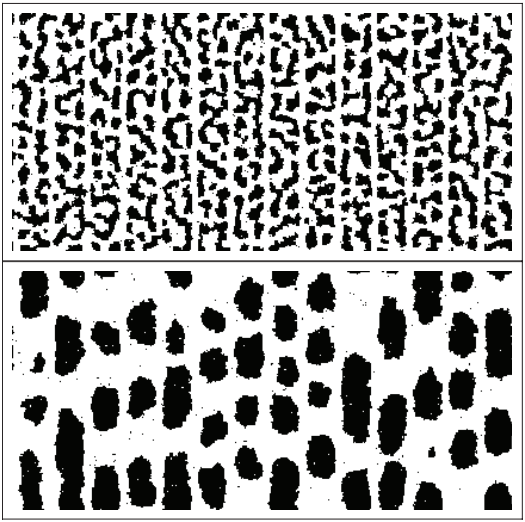}}
\caption{Two different snapshots  of   membrane domains  as obtained from canonical Monte
Carlo simulations
for area fraction $X_{o} =0.4$ after  (top) $10^5$  and (bottom) $10^7$ MC moves per patch. The black and white regions correspond to ${\cal L}_o$ and ${\cal L}_d$ patches. respectively.  \cite{footnote5}}
\label{snapshot}
\vspace*{-0.3cm}
\end{center}
\end{figure}

The horizontal dotted line in Fig.~\ref{pd_X} represents   transitions from a membrane with many striped domains to  a membrane with two  large domains. This line is given by $\phi \Delta \kappa / \lambda R = 
\phi \xi_{\rm el}  / R = 2 (2 \phi R + L)/ L$. The latter relation implies that the elastic length scale 
$ \xi_{\rm el} = \Delta \kappa / \lambda$ must be compared with the  topographical 
length scale
\be 
\xi_{\rm to} \equiv 2 R \left( 2 \phi R +L \right) / \left( \phi \, L \right) ,  
\label{xi_to}
\ee
which depends only on the substrate topography, see Fig.~\ref{cartoon}. 
For  $ \xi_{\rm el}   > \, \xi_{\rm to}  $, 
the bending rigidity contrast $\Delta \kappa$ dominates over the line tension $\lambda$ and 
the membrane contains many striped ${\cal L}_o$ and ${\cal L}_d$ domains  
as confirmed by Monte Carlo simulations, see   Fig.~\ref{snapshot}. For $\xi_{\rm el} < 
\xi_{\rm to}$, on the other hand, corresponding to large line tension and/or small bending rigidity 
contrast,  the membrane will completely  phase separate into  two large domains.

{\sl Monte Carlo simulations} -- The phase diagram obtained by energy minimization  has been confirmed by Monte Carlo (MC)  simulations. To perform such  simulations, the membrane surface is divided up into square patches of side length~$a$. This leads to a square lattice that is composed of parallel stripes of length $L_\parallel$. These stripes have  alternating   width  $L$ and $2 \phi R$, compare Fig.~\ref{cartoon}, where the stripes  with width $2 \phi R$ correspond to 
the curved ridges and valleys of the substrate surface  and will be denoted by ${\cal C}$. 
Each   patch is labeled by a pair of integer numbers $i=(i_x , i_y)$ with $1 \le i_x \le N_x$ and $1 \le i_y \le N_y$. Periodic boundary conditions are imposed in both directions.

On each   patch $i$, we place the occupation number $n_i $ with  $n_i = 0$ and 
$n_i = 1$ if the square patch   $i$ is occupied by an ${\cal L}_d$ and an  ${\cal L}_o$ domain, respectively.    The discretized  membrane  energy $\cE_{\rm dis}$ is then given by 
\be 
\frac{\cE_{\rm dis}}{a^2} =\frac{\Delta \kappa}{2 R^2} \sum_{ i \in {\cal C} } n_i  - \Delta W \sum_i n_i
+ \frac{\lambda_{0}}{2 a} \sum_{ \langle i  j \rangle } n_i (1- n_j )
\label{TotalEnergyDis}
\ee
corresponding to the bending, adhesion, and line energy   as in 
(\ref {TotalEnergy}). The first term  contains a summation over all patches $i$ that are contained 
in the curvature stripes ${\cal C}$ while  the last term contains a summation 
over all pairs of nearest neighbor patches $\langle i j \rangle$.  The coupling constant $\lambda_{0}$
is related to the  line tension $\lambda$ via  \cite{Onsager44}
$ \lambda a/T = \lambda_0 a/T  - \ln [(1+e^{- \lambda_0 a /T})/ (1-e^{- \lambda_0 a/T}) ]$
with temperature $T$   in energy units. 

For the grand-canonical ensemble, we determined the equilibrium states of  the discretized model as given by 
(\ref{TotalEnergyDis}) using Glauber dynamics \cite{BinderBook}. During each   move of this MC algorithm,  a membrane patch $i$ is chosen randomly and the value of corresponding variable $n_i$  is changed from $0$ to $1$ or from $1$ to $0$. This trial move is   then accepted according to the Metropolis criterion. 
In this way, we studied the  stability of different domain patterns   and determined  the critical nucleation size  of the domains as a function of the model parameters. Discontinuous changes in the critical nucleation size of the domains indicate transitions between the homogeneous membrane phases (I) or (II) and the striped membrane phase (III). Therefore, the domain stability analysis based on the MC simulations allows us to determine the phase boundaries where parallel stripes of the ${\cal L}_o$ and ${\cal L}_d$ phase coexist in the membrane, see Fig.~\ref{pd3}. For small lattice parameter $a$ with  $a \ll \min \{ 2 \phi R, L \}$,  our simulation results are in a good agreement with the mean field predictions (\ref{regime_iii}). The MC data included in  Fig.~\ref{pd3} were obtained for $L/a= 120$. 

We also studied the process of domain formation within the canonical ensemble, in which the    number $\mathcal{N} = \sum_{i} n_i$ of membrane patches in the ${\cal L}_o$ phase is kept constant. This constraint was incorporated via 
 diffusive Kawasaki dynamics \cite{BinderBook}, in which  two neighboring patches $i$ and $j$ with $n_i =1$ and     $n_j =0$   are chosen randomly and the values of  $n_i$ and $n_j$ are swapped according to the Metropolis criterion.
 Starting from a random initial configuration $\{ n_i \}$, 
 we then observed   anisotropic coarsening processes as illustrated in Fig.~\ref{snapshot}.  
 In this example, the model parameters 
 fulfill  the inequality (\ref{xi_to}) corresponding to  
 coexistence between phases (I) and (III). 
  As the system evolves in time, the more rigid patches start to aggregate  in the flat membrane stripes and to  form small domains of irregular shapes. Some of these domains shrink and, eventually, vanish whereas other domains grow and become elongated  parallel to the flat membrane stripes. 
 Since the diffusion-limited exchange process across the  ridges and valleys between  the flat stripes,  is rather slow, we performed
 additional simulations starting from the  state of complete phase separation and confirmed that this
 state is stable apart from shape fluctuations of the 1-dimensional interface between the coexisting phases (I)  and (III).

{\sl Summary} -- We have theoretically studied  domain patterns in multi-component  membranes supported on corrugated  substrates as  in Fig.~\ref{cartoon}. We showed that the   process of pattern formation 
is governed by the 
competition between bending rigidity contrast and line tension.  The corresponding phase diagrams are 
displayed in   Fig.~\ref{pd3}
and in  Fig.~\ref{pd_X}    for the grand-canonical and canonical ensemble, respectively. 
The stability of substrate-induced domain patterns can be understood in terms of two length 
scales, the elastic length $\xi_{\rm el} = \Delta \kappa / \lambda$ and the topographical length 
$\xi_{\rm to} $ as given by (\ref{xi_to}).  The substrate induced patterns with many 
striped membrane domains represent the true equilibrium states of the system 
provided 
$\xi_{\rm el} > \xi_{\rm to}$.   For the coexistence regions of the 
${\cal L}_o$ and ${\cal L}_d$ phases as studied experimentally, the elastic length scale 
$\xi_{\rm el} $ is of the order of 100~nm as follows from the experimental data in \cite{baum03+baum05}. 
This implies that both regimes $\xi_{\rm el} > \xi_{\rm to}$  and $\xi_{\rm el} <  \xi_{\rm to}$ 
should be experimentally accessible.


\begin{thebibliography}{99}
\bibitem{footnote2} In general, our theory applies to  any  coexistence region for  two liquid phases within the membrane.
\bibitem{footnote5} 
The snapshots shown in Fig.~\ref{snapshot} were obtained for the following parameters: 
$\phi \, \Delta \kappa /  ( \lambda_0 R) =8/3$, 
$2 \phi R /L =1/5$, $L/a=25$,  $\lambda_0 a /T=1.5$ corresponding to 
 $\lambda \simeq 0.6974 \, \lambda_0$.
\end{thebibliography}
\end{document}